\documentstyle{elsart}
\begin{document}
\begin{frontmatter}
\title{Dependence of the superconducting effective mass on doping in cuprates}
\author[physics,kolja]{N.Kristoffel\thanksref{mail},}
\author[physics]{P.Rubin\thanksref{mail}}
\address[physics]{Institute of Physics, University of Tartu,
Riia 142, 51014 Tartu, Estonia}
\address[kolja]{Institute of Theoretical Physics, University of Tartu, T\"ahe 4,
51010 Tartu, Estonia}
\thanks[mail]{Corresponding authors: Tel: +372 742 8164; fax: +372 738 3033;
E-mail: kolja@fi.tartu.ee (N.Kristoffel), rubin@fi.tartu.ee (P.Rubin)}

\begin{abstract}
Using a doping-determined multiband model spectrum of a "typical''
cuprate the effective mass of the paired carriers is calculated on
the whole doping scale. Large $m_{ab}$ values quench rapidly with
leaving the very underdoped region. Further slower diminishing of
$m_{ab}$ reproduces the trend towards restoring the Fermi-liquid
behaviour in cuprates with progressive doping. The interband
superconducting condensate density ($n_s$) shows similar behaviour
to the transition temperature and superconducting gaps. The
$n_s(0)/m_{ab}$ ratio has an expressed maximum close to optimal
doping as also the thermodynamic critical field. All the
overlapping band components are intersected by the chemical
potential at this. The pairing strength and the phase coherence
develop simultaneously. In spite of its simplicity, the model
describes the
behaviour of various cuprate characteristics on the doping scale.\\

\end{abstract}
\end{frontmatter}

\section{Introduction}
The superconducting condensate density ($n_s$) characterizes the order
parameter of this macroscopic quantum commonwealth. The way to cuprate
high-T$_c$ superconductivity includes (necessarily) a doping treatment.
The nature of the reorganizations in the physical-chemical basis of the
material and the variation of the superconducting properties on the doping
scale become of primarily interest. At present there remain nevertheless
some debatable aspects in the cuprate superconductivity until the pairing
mechanism itself. Correspondingly this concerns also the behaviour of the
superconducting condensate density on doping, e.g. [1,2].

An essential property which determines numerous applications of
superconductors is the penetration depth. It contains $n_s$ with
the superconducting condensate effective mass $m$ in the
combination $m/n_s$. The doping dependence of the
superconductivity playground CuO$_2$ plane $m_{ab}$ in cuprates is
practically unknown. Usually one supposes a constant value $m \sim
xm_0$, with $x$ between, say, 2 and 5 ($m_0$ is the free electron
mass). The main aim of the present contribution is the calculation
of $m_{ab}$ on the whole scale of the hole doping ($p$) in
association with other cuprate superconductivity characteristics.

We use a very simple, partly postulative model of a "typical'' cuprate
superconductor which uses only general knowledge on the system. The model
has been started by Ref. [3,4] and developed in [5-7]. The following comparison
of the outcome of the model for various properties with the observations
is expected to illuminate in some extent the background physics.

The model supports on the two-component scenario of cuprate superconductivity
[8,9] which states the essential functioning of the doping-created defect
subsystem besides the itinerant one. This means that the background electron
spectrum shows essential dynamics under doping. Bare normal state gaps
are assumed between the mentioned subsystems and supposed to be quenched
by progressive doping. Overlap dynamics of the bands appears now as a
novel source of critical doping concentrations. The nature of the minimal
quasiparticle excitation energies changes with doping in accordance with the
chemical potential position.  This explains
naturally the presence of pseudogaps in the model. The pseudogaps appear as
precursors on the doping scale (not on the energetic one) to the
superconducting gaps and survive for $T>T_c$ as normal state gaps.

The pairing interaction is supposed to be of interband pair transfer [10]
type between the itinerant and defect states. This mechanism is seemingly
the most effective in serving high transition temperatures in a simple
way [11,12].

The calculated phase diagram of cuprate energetic characteristics
($T_c$, superconducting- and pseudogaps, the condensation energy) [5,6]
agree qualitatively well with the experimental results. Controversial
statements in the literature on interrelations and coexistence of various
gaps in distinct doping regions have been elucidated. Compounds with two or
one pseudogap in the charge channel can described.

\section{The physical model}
We proceed with the description of the physical content of the
model background used. It enables the understanding of the
behaviour of the characteristics investigated. The cuprate
electron spectrum created and reorganized by doping is chosen as
follows. The mainly oxygen itinerant valence band ($\gamma$) top
fixes the energy zero and this band extends until $\xi =-D$. The
states in it are normalized to $1-c$, where $c$ is a measure of
the doped hole concentration ($p$). The corresponding scaling for
a given case must be made by joining a characteristic
concentration on the phase diagram. There is a huge (e.g. [13-17])
amount of appointments that doping creates new defect (midband)
states near the top of the valence band in the charge-transfer
gap. Extended doping brings them to merge with the valence band.
The functioning of the defect subsystem is anisotropic in the
momentum space [18]. At least the "hot'' ($\pi ,0$)-type and
"cold'' $\left(\frac{\pi}{2}, \frac{\pi}{2}\right)$-type regions
must be distinguished. Accordingly we introduce two defect system
subbands characterized by energy intervals $d_1-\alpha c$ and
$d_2-\beta c$, i.e. they expand down from energies at $c=0$. The
overlap of these bands with the valence band is reached at
$c_{\alpha}=d_1\alpha\sp{-1}$ and $c_{\beta}=d_2\beta\sp{-1}$.
Note that the infrared manifestations of the defect subsystem are
suppressed in favour of the free carriers (Drude peak appearance)
with progressive doping  [19]. We take $d_1$ and $d_2$ to be
positive: the optical charge-transfer gap is reduced by doping
[20]. The choice $c_{\beta}<c_{\alpha}$ accounts for that the
lowest doping-created states belong to the cold subsystem. The
weight of the defect states is taken to be $c/2$, cf. [17].

The $2D$ (CuO$_2$ planes) densities of states in the bands read
$\rho_{\alpha}=(2\alpha )\sp{-1}$, $\rho_{\beta}=(2\beta )\sp{-1}$,
$\rho_{\gamma}=(1-c)D\sp{-1}$. There are the following different
arrangements of the bands and the chemical potential ($\mu$). At very
underdoping $c<c_{\beta}$, $\mu_1=d_2-\beta c$ is connected with the
"cold'' $\beta$-band and charge carriers become concentrated here, cf. [13].
For $c>c_{\beta}$, $\mu_2=(d_2-\beta c)[1+2\beta (1-c)D\sp{-1}]\sp{-1}$
intersects both ($\beta ,\gamma$)-bands. The Fermi level shifts into the
valence band, cf. [21]. The overlap of the narrow defect $\beta$-band
with the itinerant band leads to the formation of two sheets of the
Fermi surface. The one at $\left(\frac{\pi}{2}, \frac{\pi}{2}\right)$ is
holelike with the dominating itinerant contribution and the other with a
tendency to form an electronlike "flat band'' with lowering $\mu$. This
behaviour is  in agreement with the observation of two Fermi-sheets [22]
and of a single "hole barrel'' at $\left(\frac{\pi}{2}, \frac{\pi}{2}\right)$
[23] and a flatband at ($\pi ,0$) [24].

For the expressed dopings larger than $c_0$, determined by $d_1-\alpha c_0=
\mu_2$, the role of the ($\pi ,0$)-type region increases essentially as
also Refs. [25,26] state. Now $\mu_3=[\alpha d_2+\beta d_1-2\alpha\beta c]
[\alpha +\beta +(1-c)2\alpha\beta D\sp{-1}]\sp{-1}$ intersects all
three overlapping bands. For extended overdoping $c>c_1$, where $c_1$ is
defined by $d_2-\beta c=\mu_3$, the chemical potential $\mu_4=(d_1-\alpha c)
[1+2\alpha (1-c)D\sp{-1}]\sp{-1}$ falls out of the cold defect band. The
chemical potential is not affected significantly by pairing and its general
trend agrees with the results of special investigations [27-29].

\section{Necessary formulas}
Our basic Hamiltonian with the coupling ($W$) of itinerant and defect
subsystems by the pair-transfer interaction reads
\begin{equation}
H=\sum_{\sigma ,\vec{k},s}\epsilon_{\sigma}(\vec{k})
a\sp +_{\sigma ,\vec{k},s}a_{\sigma ,\vec{k},s}+
W\sum_{\sigma ,\sigma '}{}'\sum_{\vec{k},\vec{k}'}\sum_{\vec{q}}
a\sp +_{\sigma\vec{k}\uparrow}a\sp +_{\sigma (-\vec{k}+\vec{q})\downarrow}
a_{\sigma '(-\vec{k}'+\vec{q})\downarrow}
a_{\sigma '\vec{k}'\uparrow}\; .
\end{equation}

Here $\epsilon_{\alpha}=\xi_{\sigma}-\mu$, $s$ is the spin index,
$\sigma$ counts the bands and $\vec{q}$ is the pair momentum with the
components from the same bands. Various aspects of the work with (1) at
$\vec{q}=0$ for calculating the superconductivity energetic characteristics
can be followed in [3,5,10]. The superconductivity gap
parameters are defined as
\begin{eqnarray}
\Delta_{\gamma} & = & 2W\sum_{\vec{k},\tau}{}\sp{\tau}
<a_{\tau \vec{k}\uparrow}a_{\tau -\vec{k}\downarrow}>\; ,\\ \nonumber
\Delta_{\tau} & = & 2W\sum_{\vec{k}}
<a_{\sigma -\vec{k}\downarrow}a_{\sigma \vec{k}\uparrow}>\; ,
\end{eqnarray}
where $\sum {}\sp{\tau}$ means the integration with the densities
$\rho_{\alpha ,\beta}$ over the corresponding energy intervals of the defect
system subbands ($\tau =\alpha$, $\beta$; $\Delta_{\alpha}=\Delta_{\beta}$).
The nongapped nature of the cold subsystem [18] can be accounted by the
multiplication of $\Delta_{\alpha}$ with the suitable $d$-symmetry factor.

The gap equation reads ($\theta =k_BT$)
\begin{eqnarray}
\Delta_{\sigma}=W\sum_{\vec{k},\tau}{}\sp{\tau}\Delta_{\tau}(\vec{k})
E_{\tau}\sp{-1}(\vec{k}) th\frac{E_{\tau}(\vec{k})}{2\theta}\\
\nonumber \Delta_{\tau}=W\sum_{\vec{k}}\Delta_{\gamma}(\vec{k})
E_{\gamma}\sp{-1}(\vec{k}) th\frac{E_{\gamma}(\vec{k})}{2\theta}
\end{eqnarray}
with the usual form of the quasiparticle energies
$E_{\sigma}(\vec{k})=\sqrt{\epsilon_{\sigma}\sp 2(\vec{k})
+\Delta_{\sigma}\sp 2(\vec{k})}$. The density of the paired carriers is
\begin{equation}
n_{s}=\frac{1}{2}\left[ \sum_{\vec{k}}\frac{\Delta\sp
2_{\sigma}(\vec{k})} {E\sp 2_{\sigma}(\vec{k})} th\sp
2\frac{E_{\gamma}(\vec{k})}{2\theta}+
\sum_{\vec{k}}{}\sp{\tau}\frac{\Delta\sp 2_{\tau}(\vec{k})} {E\sp
2_{\tau}(\vec{k})} th\sp 2\frac{E_{\tau}(\vec{k})}{2\theta}
\right]\; .
\end{equation}

The free energy corresponding to the Hamiltonian (1) has been calculated in
[30] (see also [7]), where the paired carrier effective mass isotope
defect has been investigated. The "soft'' order parameter [30] with the
critical behaviour at  $T_c$ is characterized by the effective mass
\begin{equation}
m_{ab}=\frac{1}{2} \frac{(\eta_{\alpha}+\eta_{\beta}+\eta_{\gamma})
(\delta_{\alpha}+\delta_{\beta}+\delta_{\gamma})}
{(\eta_{\alpha}+\eta_{\beta})\delta_{\gamma}m_{\gamma}\sp{-1}+
\eta_{\gamma}(\delta_{\alpha}m_{\alpha}\sp{-1}+\delta_{\beta}m_{\beta}\sp{-1})}\; ,
\end{equation}
where $m_{\sigma}=2\pi h\sp 2\rho_{\sigma}V\sp{-1}$, and $V=a\sp 2$ for
the CuO$_2$ plaquette. For $\mu -s$ being not too close to limiting
energies $\Gamma_{0\sigma}$ and $\Gamma_{c\sigma}$ of the bands the
following formulas for the quantities entering (5) can be used
\begin{equation}
\eta_{\sigma}=W\rho_{\sigma}\ln \left[\left( \frac{2\gamma}{\pi}\right)\sp 2
\theta_c\sp{-2}| \Gamma_{0\sigma}-\mu||\Gamma_{c\sigma}-\mu|\right] \; ,
\end{equation}
\begin{equation}
\delta_{\sigma}=\frac{7}{2}\zeta (3)W\rho_{\sigma}|\mu -\Gamma_{0\sigma}|
(\pi\theta_c)\sp{-2}\; ,
\end{equation}
when $\mu$ is located in the integration region ($\zeta (x)$ is the
zeta-function; $\gamma =\exp (0.577)$).

If $\mu$ lies out of the band $\delta_{\sigma}=0$ and
\begin{equation}
\eta_{\sigma}=W\rho_{\sigma}\ln \left|
\frac{\Gamma_{c\sigma}-\mu}{\Gamma_{0\sigma}-\mu}\right|\; .
\end{equation}

According to (5) the supercarrier effective mass depends on the
position of the spectral
components and the chemical potential at a given doping.  The
mixed nature of the excitations in the multiband model is reflected in
the expression (5). With the approximations used $m_{ab}$ is temperature
independent.

\section{Calculated doping dependences}
The calculations of $T_c$, superconductivity gaps, etc. have been made
by numerical integration using a plausible parameter set of Ref. [5]. The
doping dependences of the cuprate energetic characteristics are illustrated
in [5,6]. Here we represent the zero-temperature superfluid density and the
condensation energy in Fig.1 {\it vs.} hole doping $p=0.28c$ (it has been
taken $p=0.16$ for the $T_c(max)$). The condensation energy is represented by
the thermodynamic critical field as
\begin{equation}
H_{c0}=\sqrt{4\pi [(\rho_{\alpha}+\rho_{\beta})\Delta_{\alpha}\sp 2+
\rho_{\gamma}\Delta_{\gamma}\sp 2]}\; .
\end{equation}
The bell-like curves of $T_c$, $\Delta_{\sigma}$ [5,6], $n_s$ and $H_{c0}$
show the similar behaviour in agreement with the results of [2,31]. Our
calculation of the $\xi_{ab}$ correlation length {\it vs} $p$ has given a
valley-profile like curve [7]. The second critical field $H_{c2}(0)$
calculated from $\xi_{ab}$ has also a well expressed maximum [7]. In this
manner the strength of the pairing and the phase coherence develop and
vanish simultaneously in our model. Analogous conclusion has been done in
a number  of recent investigations [1,2,32-35]. The $T_c$ maximum peak on
the doping scale is a result of the electron spectrum doping-driven dynamics
which brings all the band components into overlap at the Fermi energy.
Bands overlap dynamics appears as a novel source of critical doping
concentrations. In the normal state at $c_{\beta}$ and $c_0$ insulator-to-metal
(in the cold and hot subsystems, respectively) are expected to appear. This is
in agreement with the observations [2,36,37]. At $c_0$ also the large
pseudogap vanishes, cf. [1,2], due to this overlap. The following overdoped
regime corresponds to higher carrier concentrations but to smaller
superconducting carrier concentrations. The scatterings which cause pairing
by the interband mechanism are reduced here.

It must be also mentioned that the bare itinerant -- defect gaps are not
manifested in the superconducting density curve because the interband
nature of the pairing.

The $n_s(T)$ dependence is illustrated in Fig.2 and describes also the
penetration depth
\begin{equation}
\lambda =\left[ \frac{x m_0c\sp 2a\sp 2l}{4\pi
e^2n_s(T)}\right]\sp{1/2}
\end{equation}
dependence on temperature. Here $m_{ab}$ is expressed through the free
electron mass $m_0$ as $m_{ab}=xm_0$ and $l$ is the c-axis lattice constant.

The doping-dependence curve of the paired carrier effective mass
is given in Fig.3. To the authors knowledge it has not been
obtained earlier. It is seen that $m_{ab}$ cannot be assumed to be
constant on the whole doping scale. However, in the actual region
of remarkable $T_c$ values, a rough estimation near $x\sim 3$, as
often supposed, can be considered as acceptable for estimations.
It must be stated that the parameter set used can serve only
plausible estimations on the quantitative level.

The large values of $x$ at very underdoping correspond to the large
effective mass of the narrowest defect $\beta$-band. Starting from
$c_{\beta}$ the contribution of the wide valence band carriers is
continuously added. The kink at $c_0$ corresponds to the simultaneous
action of the $\alpha -\beta$ carriers. The essential decrease in $m_{ab}$
after reaching $c_1$ is connected with the vanished contribution of the
heaviest $\beta$-carriers. The superconducting carrier effective mass
reflects the structure of the electron spectrum and reproduces the well known
trend of the superconducting collective towards the normal Fermi liquid
behaviour with doping in cuprates.

The essential ratio $n_s(0)/x$ determining the inverse penetration
depth squared is shown {\it vs} doping in Fig.4. The superfluid
density bell-like dependence dominates over the effective mass
changes. As the result, $\lambda\sp{-2}$ is characterized by a
curve with a well expressed maximum on the doping scale. This is
in agreement with the experimental findings [1,2,32]. The peaked
behaviour of $n_s(0)/m_{ab}$ is a natural consequence of the
interband superconductivity. This answers the question rised in
Ref. [1] about the decrease of $n_s$ at overdoping. The sublinear
$T_c$ {\it vs} $n_s$  (or $\lambda\sp{-2}$) plot at underdoping,
which has been observed in a number of investigations [1,38], is
reproduced also by the present model -- Fig.5.

Having in mind the results obtained for the energetic
characteristics of cuprates [3-6] and the coherence length [7],
the results of the present work, and also for the effect of
photodoping [39], one can conclude that the model under
consideration, despite of its simplicity, is able to describe the
cuprate charge-channel associated properties on the whole doping
scale.

The spin-channel effects are out of the scope of the model at present.
However, there seems to be some correspondence with the models like [40]
where the magnetic properties become explained. The ability of the authors
model to reproduce qualitatively the behaviour of various cuprate
superconducting characteristics on doping is expected to stimulate to
fill it in with precised suppositions and quantitative refinements.

\section*{Acknowledgement}

This work was supported by Estonian Science Foundation grant No 6540.
\section*{References}
\begin{itemize}
\item[[1]]  C. Bernhard et al.,  Phys. Rev. Lett. {\bf 86} (2001) 1614.
\item[[2]]  J.L. Tallon et al.,  Phys. Rev. B {\bf 68} (2003) 180501(R).
\item[[3]]  N. Kristoffel, P.Rubin, Physica C {\bf 356} (2001) 171;
Solid State Commun. {\bf 122} (2002) 265.
\item[[4]]  N. Kristoffel, P. Rubin, Eur. Phys. J. B {\bf 30} (2002) 495.
\item[[5]]  N. Kristoffel, P. Rubin, Physica C {\bf 402} (2004) 257.
\item[[6]]  N. Kristoffel, P. Rubin, Proc. Estonian Acad. Sci. Phys. Math.
 {\bf 54} (2005) 98; cond.-mat./0408574v1 (2004).
\item[[7]]  N. Kristoffel, T. \"Ord, P. Rubin, cond.-mat./0504431v1 (2005).
\item[[8]]  K.A. M\"uller, Physica C {\bf 341-348} (2000) 11.
\item[[9]]  D. Mihailovic, K.A. M\"uller, in  Proceedings of the
NATO ASI Materials Aspects of High-T$_c$ Superconductivity, Kluwer,
Dordrecht, 1997, p.1.
\item[[10]] N. Kristoffel, P. Konsin, T. \"Ord, Riv. Nuovo Cim. {\bf 17}
(1994) 1.
\item[[11]] H. Suhl, B.T. Matthias, L.R. Walker, Phys. Rev. Lett.
{\bf 3} (1959) 552.
\item[[12]] V.A. Moskalenko, Fiz. Met. Metalloved. {\bf 8} (1959) 503.
\item[[13]] A. Ino et al., Phys. Rev. B {\bf 65} (2002) 094504.
\item[[14]] Y. Ando et al., Phys. Rev. Lett. {\bf 87} (2001) 017001.
\item[[15]] H. Ihara, Physica C {\bf 364-365} (2001) 289.
\item[[16]] C.C. Homes et al. Phys. Rev. B {\bf 67} (2003) 184516.
\item[[17]] A.A. Borisov , V. A. Gavrichkov, S. G. Ovchinnikov, Modern Phys. Lett. B {\bf 17} (2003) 479.
\item[[18]] T. Timusk, B. Statt, Rep. Progr. Phys. {\bf 62} (1999) 61.
\item[[19]] P. Calvani, phys. stat. sol. b {\bf 237} (2003) 194.
\item[[20]] Y. G. Zhao et al., Phys. Rev. B {\bf 63} (2001) 132507.
\item[[21]] H. Romberg et al., Phys. Rev. B {\bf 42} (1990) 8768.
\item[[22]] P.V. Bogdanov et al., Phys. Rev. B {\bf 64} (2001) 180505.
\item[[23]] J. Mesot et al., Phys. Rev. B {\bf 63} (2001) 224516.
\item[[24]] Y. Yoshida et al., Phys. Rev. B {\bf 63} (2001) 220501.
\item[[25]] A.D. Gromko et al., Phys. Rev. B {\bf 68} (2003) 174520.
\item[[26]] S. Sugai et al., Phys. Rev. B {\bf 68} (2003) 184504.
\item[[27]] A. Ino et al., Phys. Rev. Lett. {\bf 79} (1997) 2101.
\item[[28]] T. Tohyama, S. Maekawa, Phys. Rev. B {\bf 67} (2003) 092509.
\item[[29]] N. Harima et al., Phys. Rev. B {\bf 67} (2003) 172501.
\item[[30]] T. \"Ord, N. Kristoffel, phys. stat. sol. b {\bf 216} (1999) 1049.
\item[[31]] T. Schibauchi et al., Phys. Rev. Lett. {\bf 86} (2001) 5763.
\item[[32]] D.L. Feng et al., Science {\bf 289} (2000) 277.
\item[[33]] T. Schneider, Physica B {\bf 326} (2003) 289.
\item[[34]] M.R. Trunin, Yu. A. Nefyodov, A.F. Shevchun, Phys. Rev. Lett.
{\bf 92} (2004) 067006.
\item[[35]] R. H. He et al., Phys. Rev. B {\bf 69} (2004) 220502(R)
\item[[36]] F. Venturini et al., Phys. Rev. Lett. {\bf 89} (2002) 107003.
\item[[37]] X.F. Sun, K. Segawa, Y. Ando, Phys. Rev. Lett. {\bf 93} (2004)
107001.
\item[[38]] Y.J. Uemura et al., Phys. Rev. Lett. {\bf 66} (1991) 2665.
\item[[39]] N. Kristoffel, P. Rubin, Physica C {\bf 418} (2005) 49.
\item[[40]] A.S. Alexandrov, P.P. Edwards, Physica C {\bf 331} (2000) 97.
\end{itemize}

\newpage
Figure captions \\ \\

Fig. 1. The superfluid density (full line) and the thermodynamic
critical field (dashed line) on the doping scale.

Fig. 2. The dependence of the superconducting density on temperature
for the doping $p=0.14$.

Fig. 3. The supercarrier effective mass dependence on doping.

Fig. 4. The inverse penetration depth squared {\it vs} doping representated
by $n_s(0)/x$; $m_{ab}=xm_0$.

Fig. 5. The Uemura type plot.

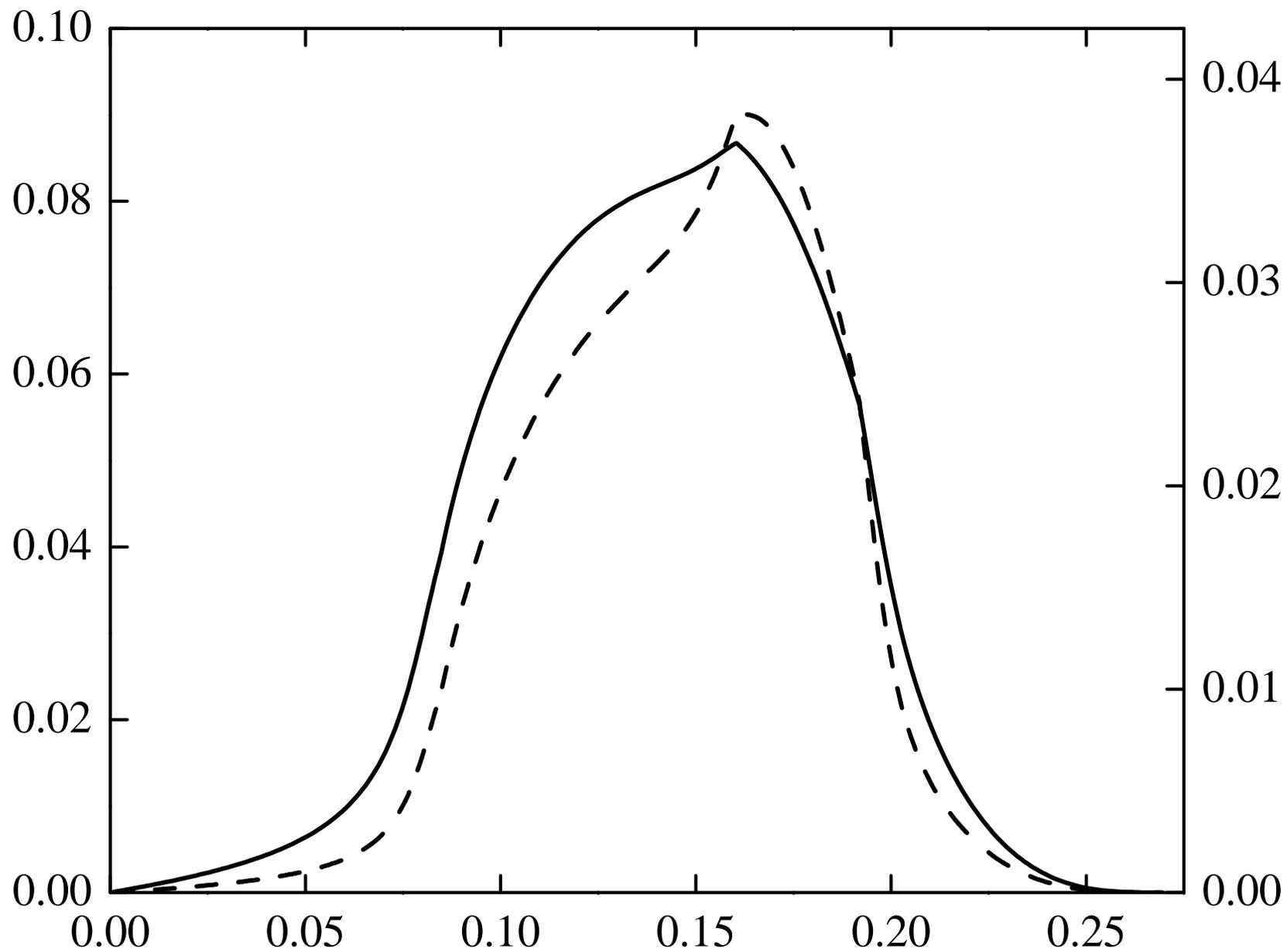
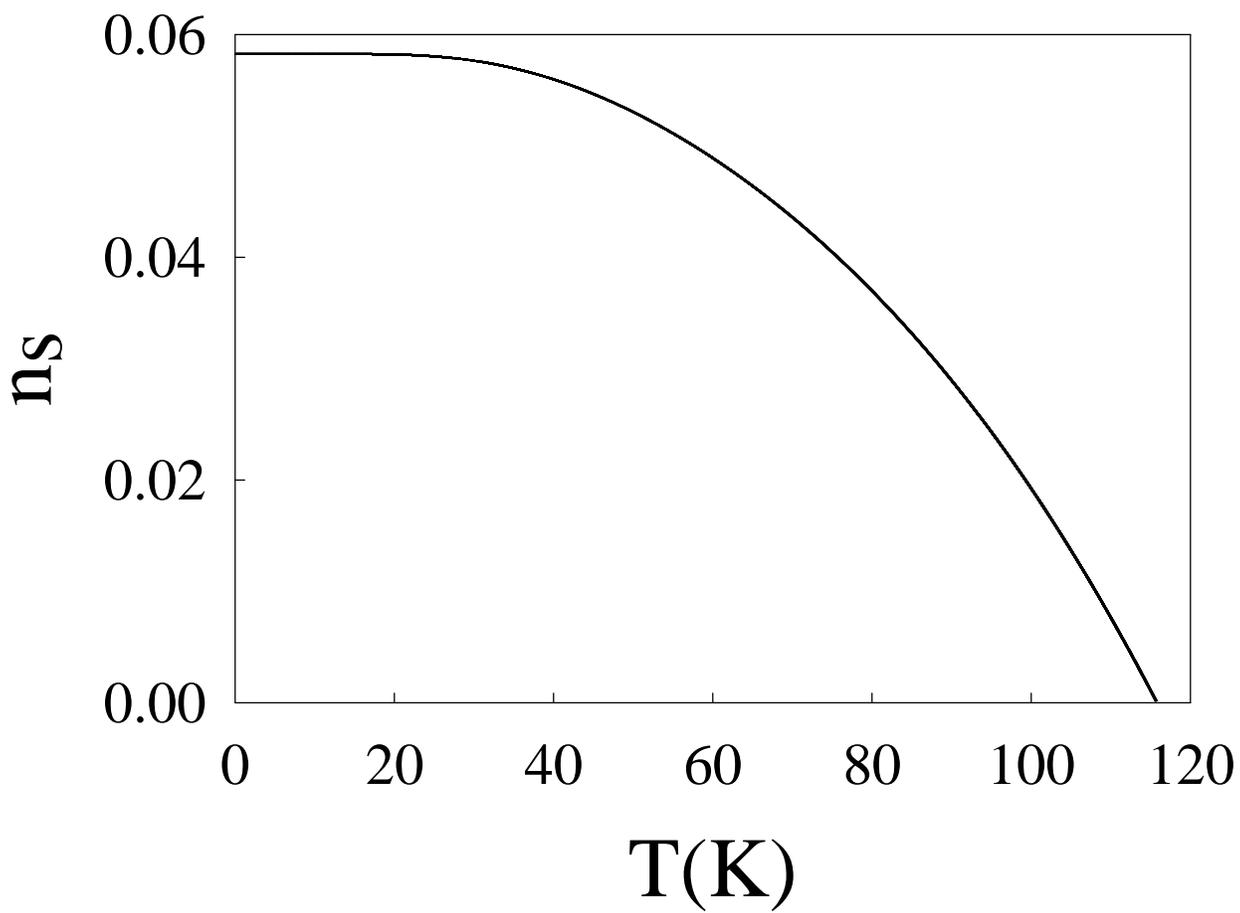
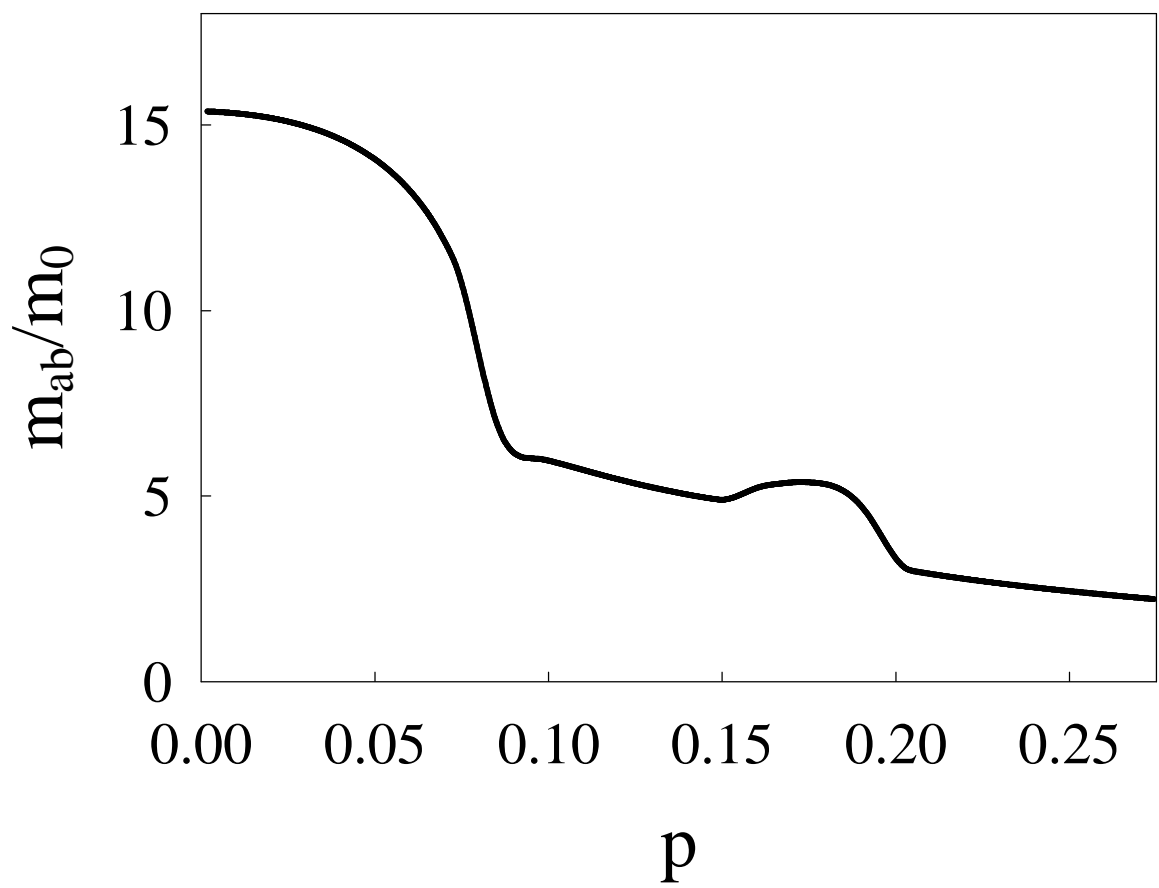
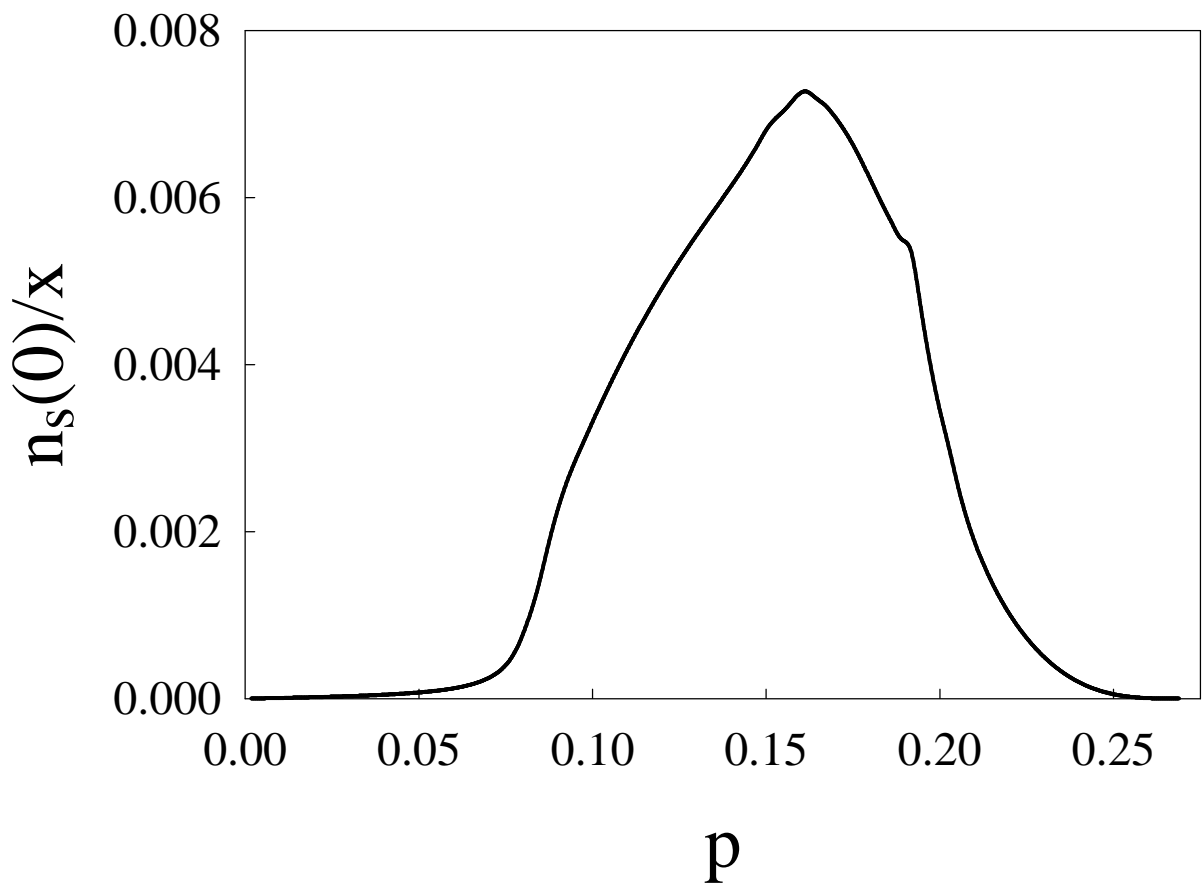
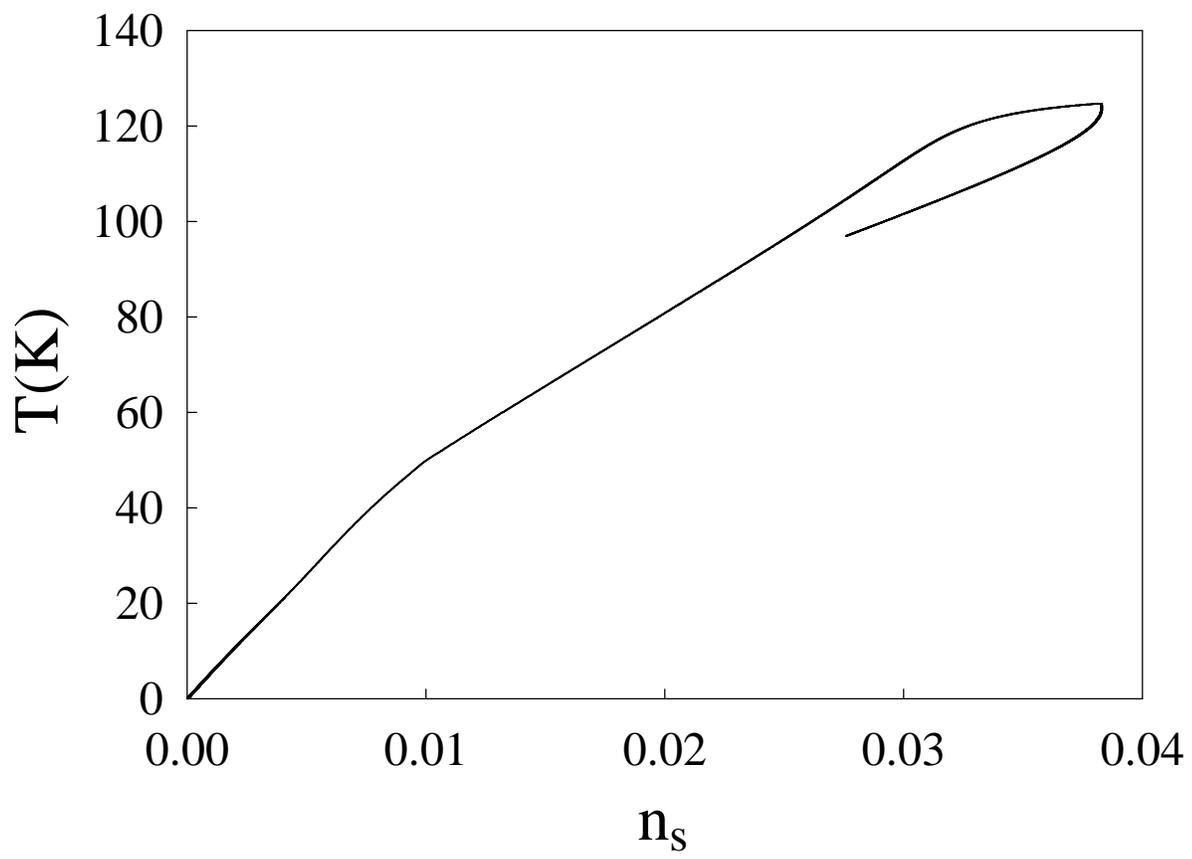
\end{document}